\documentclass[twocolumn]{aastex62}

\usepackage[normalem]{ulem}
\usepackage{footnote}
\usepackage{subfigure}
\usepackage{xspace}
\usepackage[utf8]{inputenc}
\usepackage{multirow,amsmath}
\usepackage{float}
\usepackage{apjfonts}

\newcommand\msun{{\rm M}_{\odot}}                   
\newcommand\rsun{{\rm R}_{\odot}}                   
\newcommand\lsun{{\rm L}_{\odot}}                   
\newcommand\jsun{{\rm J}_{\odot}}                   
\newcommand{\Tc}{T_{\rm c}}                         
\newcommand{\rhoc}{\rho_{\rm c}}                    
\newcommand\rot{\Omega/\Omega_{{\rm crit}}}         
\newcommand{\BV}{Brunt-V\"{a}is\"{a}l\"{a}}         

\newcommand{\chir}{\chi_{\rho}}

\newcommand{\code}[1]{\texttt{#1}}
\newcommand{\mesa}{\code{MESA}}
\newcommand{\MESA}{\mesa}
\newcommand{\gyre}{\code{GYRE}}
\newcommand{\GYRE}{\gyre}

\newcommand{\kic}{KIC\,08626021\xspace}

\begin{document}

\title{THE IMPACT OF WHITE DWARF LUMINOSITY PROFILES ON OSCILLATION FREQUENCIES}

\author[0000-0002-0474-159X]{F.~X.~Timmes}
\affiliation{School of Earth and Space Exploration, Arizona State University, Tempe, AZ, USA}
\affiliation{Joint Institute for Nuclear Astrophysics - Center for the Evolution of the Elements, USA}

\author[0000-0002-2522-8605]{Richard H. D. Townsend}
\affiliation{Department of Astronomy, University of Wisconsin-Madison, Madison, WI 53706, USA}

\author[0000-0002-4791-6724]{Evan B. Bauer}
\affiliation{Department of Physics, University of California, Santa Barbara, CA 93106, USA}

\author[0000-0002-8107-118X]{Anne Thoul}
\affiliation{Space sciences, Technologies and Astrophysics Research (STAR) Institute, Universit\'e de Li\`ege, All\'ee du 6 Ao$\hat{u}$t 19C, Bat. B5C, 4000 Li\`ege, Belgium}

\author[0000-0002-8925-057X]{C.~E.~Fields}
\affiliation{Department of Physics and Astronomy, Michigan State University, East Lansing, MI 48824, USA}
\affiliation{Center for Theoretical Astrophysics, Los Alamos National Lab, Los Alamos, NM 87545, USA}
\affiliation{Joint Institute for Nuclear Astrophysics - Center for the Evolution of the Elements, USA}

\author[0000-0002-6828-0630]{William~M.~Wolf}
\affiliation{School of Earth and Space Exploration, Arizona State University, Tempe, AZ, USA}


\shorttitle{White Dwarf Luminosity Profiles}
\shortauthors{Timmes et al.} 

\begin{abstract}

\kic is a pulsating DB white dwarf of considerable recent interest,
and first of its class to be extensively monitored by {\it Kepler} for
its pulsation properties.  Fitting the observed oscillation
frequencies of \kic to a model can yield insights into its
otherwise-hidden internal structure.  Template-based white dwarf
models choose a luminosity profile where the luminosity is
proportional to the enclosed mass, $L_r~\propto~M_r$, independent of
the effective temperature $T_{\rm eff}$.  Evolutionary models of young
white dwarfs with $T_{\rm eff} \gtrsim$ 25,000 K suggest neutrino
emission gives rise to luminosity profiles with
$L_r$~$\not\propto$~$M_r$.  We explore this contrast by comparing the
oscillation frequencies between two nearly identical white dwarf
models: one with an enforced $L_r \propto M_r$ luminosity profile and
the other with a luminosity profile determined by the star's previous
evolution history.  We find the low order g-mode frequencies differ by
up to $\simeq$\,70\,$\mu$Hz over the range of {\it Kepler}
observations for \kic.  This suggests that by neglecting the proper
thermal structure of the star (e.g., accounting for the effect of
plasmon neutrino losses), the model frequencies calculated by using an
$L_r \propto M_r$ profile may have uncorrected, effectively-random
errors at the level of tens of ${\rm \mu Hz}$. 
A mean frequency difference of $30\,{\rm \mu Hz}$,
based on linearly extrapolating published results, suggests a template model 
uncertainty in the fit precision of
$\simeq$\,12\% in white dwarf mass, 
$\simeq$\,9\% in the radius, and
$\simeq$\,3\% in the central oxygen mass fraction.

\end{abstract}


\section{Introduction}
\label{sec:introduction}

White dwarfs (WDs) are the final evolutionary state of stars whose
zero age main-sequence (ZAMS) mass is $\lesssim 8\,\msun$
\citep{liebert_1980_aa,fontaine_2001_aa, hansen_2004_aa}, which for a
Salpeter initial mass function is $\simeq 98\%$ of stars in the Milky Way
\citep[e.g.,][]{salpeter_1955_aa,scalo_1986_aa,maschberger_2013_aa}.
The interiors of WDs encapsulate their stellar evolution history,
especially the nuclear reactions and mixing that take place during the helium burning stage
\citep{metcalfe_2002_aa, metcalfe_2005_aa, fields_2016_aa, Geronimo17, Geronimo18} 
and the initial cooling that takes place when the WD is newly born. 
The pulsation properties of variable WDs
are sensitive to their mechanical and thermal structure, and hence
asteroseismology offers the potential to probe the interior structure
and prior evolution history \citep{kawaler_1985_aa, brassard_1992_aa, 
fontaine_2008_aa, winget_2008_aa, aerts_2010_aa, althaus_2010_aa, romero_2012_aa, romero_2017_aa}.

\kic is a pulsating, He I line dominated WD belonging to the DBV and
V777 Her classes \citep{winget_1982_aa} and the first to be
extensively monitored by {\it Kepler} for its pulsation properties
\citep{ostensen_2011_aa}.  \kic shows a frequency spectrum composed of
non-radial, low order g-modes, which are sensitive to the interior
stratifications of the WD.  \citet{bischoff-kim_2011_aa} identified
seven oscillation modes from 36 months of {\it Kepler} photometric
data, some with triplet and doublet structures, to identify the
spherical harmonic $\ell$ and $m$ of several modes \citep[also
  see][]{zong_2016_aa}.  Fitting these modes to ab initio WD models
\citep[e.g., WDEC,][]{bischoff-kim_2018_aa}, they found an effective
temperature $T_{\rm eff}=29,650~{\rm K}$, mass $M=0.55\,\msun$ and
evidence for a thin He layer.  \cite{giammichele_2016_aa,
  giammichele_2017_aa, giammichele_2017_ab} pioneered new techniques
to fit observed pulsation frequencies by using flexible, parameterized
WD template models, and \citet{giammichele_2018_aa} combined
eight oscillation modes with a template model to infer \kic has a
large oxygen-dominated interior region.

\begin{figure*}[!htb]
\centering{\includegraphics{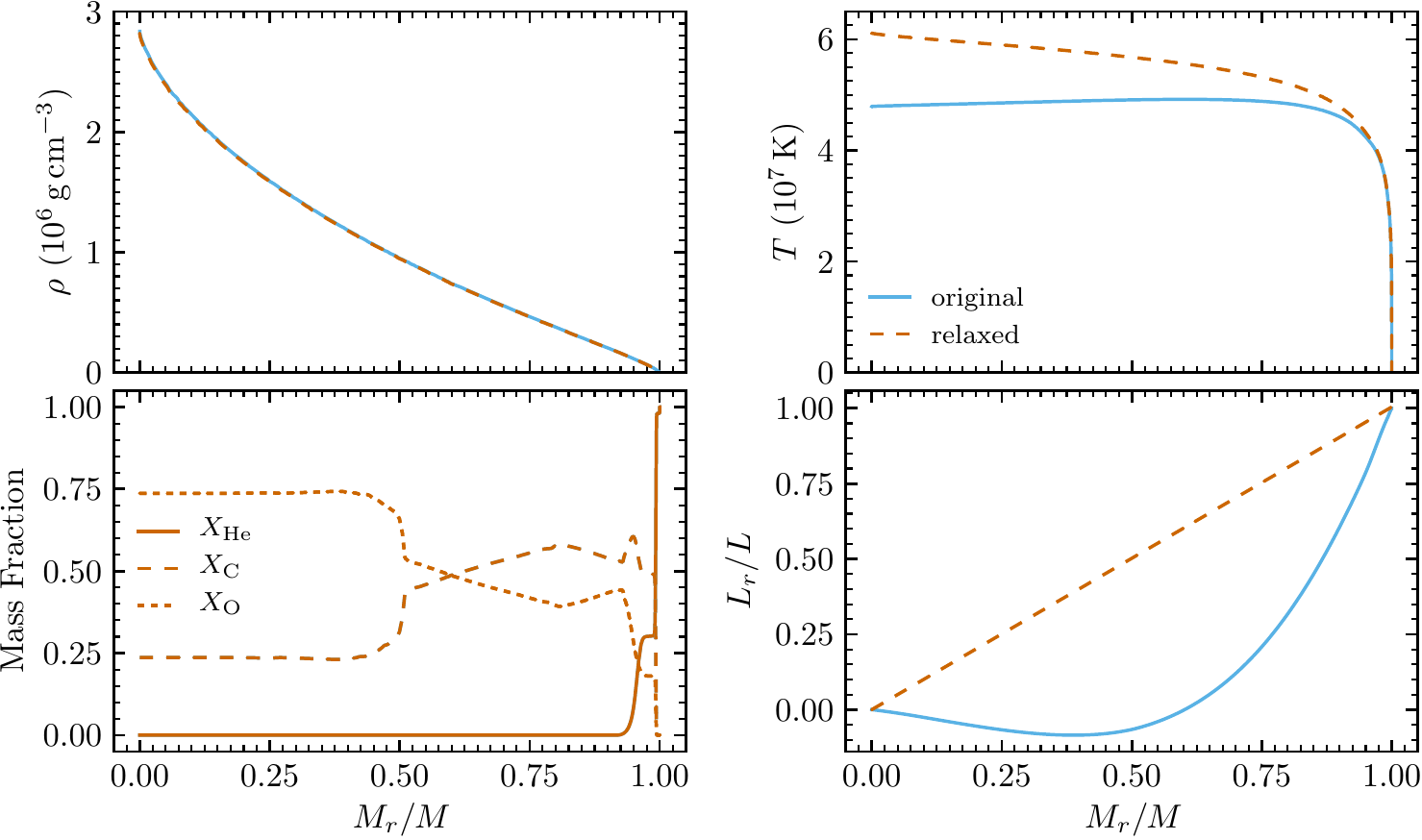}}
\caption{ Comparison of the density, temperature, mass fraction, and
  luminosity profiles between the original and relaxed \MESA\ models.
  The relaxed model is slightly hotter in the core, but density
  profiles are very similar. The mass fraction profile is identical by
  construction. The relaxed model has the imposed $L_r \propto M_r$
  profile, which is significantly different than the luminosity
  profile of the original model.  }
\label{fig:structure}
\end{figure*}

Template models typically use a WD luminosity profile $L_r \propto
M_r$, which usually assumes the WD has largely forgotten its previous
evolution history. On the other hand, stellar evolution models of
young WDs with $T_{\rm eff} \gtrsim 25,000\,{\rm K}$ suggest neutrino
emission dominates the energy loss budget for average-mass
carbon-oxygen (CO) WDs, which yields luminosity profiles with $L_r
\not\propto M_r$ \citep[e.g.,][]{vila_1966_aa, kutter_1969_aa,
  winget_2004_aa, bischoff-kim_2018_aa}.  In this Letter we explore
the difference this causes in the low order g-mode oscillation
frequencies by comparing two nearly identical WD models: one model has
an enforced $L_r \propto M_r$ luminosity profile and the other model
has a luminosity profile determined by the star's evolution history.

In Section~\ref{sec:vanilla} we present a \MESA\ model aiming towards
\kic.  In Section~\ref{sec:relaxed} we relax this model to have a $L_r
\propto M_r$ luminosity profile while keeping other characteristics
unchanged. In Section~\ref{sec:pulse} we compare the
\GYRE\ oscillation frequencies and weight functions of the two models, in
Section~\ref{sec:cooling} we discuss cooling of the WD \MESA\ model,
and we discuss the implications of our findings in Section~\ref{sec:conclude}.

\section{An Evolution Model Aiming At \kic}
\label{sec:vanilla}

We use release 10398 of the \MESA\ software instrument
\citep{paxton_2011_aa,paxton_2013_aa,paxton_2015_aa,paxton_2018_aa},
together with the published set of controls from
\citet{farmer_2016_aa} and \citet{fields_2018_aa}, to evolve a
$2.10\,\msun$, $Z=0.02$ metallicity model from the zero-age main
sequence (ZAMS) through $\simeq 12$ thermal pulses on the asymptotic
giant branch (AGB) to a CO WD.  The model includes rotation (solid
body $\rot=1.9\times 10^{-4}$ applied at ZAMS), wind mass loss
(Reimers on the red giant branch with a scaling factor of 0.1 and
Bl\"ocker on the AGB with a scaling factor of 0.5), and a 49-isotope
reaction network.  After winds have reduced the hydrogen envelope mass
to $0.01\,\msun$, we strip all of the remaining hydrogen from the
surface to form a young DB WD, which we then cool until
$L=0.137~\lsun$, matching the luminosity inferred for \kic
\citep{giammichele_2018_aa}.  Element diffusion is enabled during WD
cooling, resulting in a pure He atmosphere and smooth interior
composition transitions.

The final WD model has a mass $M=0.56\,\msun$, radius
$R=0.014\,\rsun$, effective temperature $T_{\rm eff} = 29,765\,{\rm
  K}$, surface gravity $\log g=7.90$, angular momentum $J =
1/380\,\jsun$, central density $\rhoc = 2.8 \times 10^6\,{\rm
  g\,cm^{-3}}$, central temperature $\Tc = 4.8 \times 10^7\,{\rm K}$,
central $^{16}$O mass fraction $X_{16}=0.74$, central $^{22}$Ne mass
fraction $X_{22}=0.02$, and a location where the core transitions from
being $^{16}$O dominated to $^{12}$C dominated of $M_{\rm
  trans}=0.28\,\msun$. Other than $X_{16}$, $X_{22}$, and $M_{\rm
  trans}$, this model shares many of the scalar properties with those
derived for \kic \citep{giammichele_2018_aa}.  This WD model is
referred to below as ``original model''.  Our inlist, profile, and
history files are available at {\url{http://mesastar.org}}.

\section{Imposing A Luminosity Profile}
\label{sec:relaxed}

Rather than evolve a stellar model from the pre-main sequence to a WD,
or evolve a hot initially polytropic model to a WD
\citep[e.g.,][]{bischoff-kim_2014_aa}, it can be convenient to assume
the WD has forgotten its previous evolution history by assigning a
specific profile. An example is prescribing the luminosity profile
$L_r \propto M_r$.

To facilitate comparing the pulsation properties of an evolutionary WD
model with a $L_r \propto M_r$ WD model, we modify the original model
to achieve a $L_r \propto M_r$ profile.  The energy conservation
equation
\begin{equation}
  \frac{{\rm d}L_r}{{\rm d}M_r} = \epsilon_{\rm nuc} + \epsilon_{\rm grav}
\label{eq:dldm}
\end{equation}
guarantees that a model in thermal equilibrium ($\epsilon_{\rm grav} =
0$) with a constant energy generation rate $\epsilon_{\rm nuc}$ will
satisfy $L_r \propto M_r$ (with $\epsilon_{\rm nuc}$ setting the
constant of proportionality). Thus, we replace the usual nuclear
energy generation calculations with a constant $\epsilon_{\rm nuc}$
throughout the model.  We run \MESA\ until this model relaxes to
hydrostatic and thermal equilibrium; during this relaxation process,
we allow no changes to the abundance profile (due, e.g., to diffusion
or burning).  We then compare $T_{\rm eff}$ against the effective
temperature of the original model, and iterate on $\epsilon_{\rm nuc}$
until the two match.

This process typically leads to a surface gravity different than that
of the original model. Therefore, we add a second iteration where we
adjust the WD mass $M$ while holding abundance profiles fixed as a
function of fractional mass coordinate $M_{r}/M$. Iterating on both
$M$ and $\epsilon_{\rm nuc}$ simultaneously, we obtain a model with
$L_{r} \propto M_{r}$ that agrees with the $T_{\rm eff}$ and $\log g$
of the original model to better than 0.001\,\%. This model, with
$\epsilon_{\rm nuc} = 0.471\,{\rm erg\,g^{-1}\,s^{-1}}$ and $M =
0.564\,\msun$, can be regarded as a `spectroscopic twin' to the
original model, because it shares the same effective temperature,
surface gravity, and abundances. Our inlist for creating this model is
available at {\url{http://mesastar.org}}.

Figure~\ref{fig:structure} compares the density, temperature,
abundance, and luminosity profiles of the original and relaxed
models. The density profiles are very similar, while the abundance
profiles are identical. The relaxed model follows the desired $L_{r}
\propto M_{r}$ profile, with a larger central temperature (and
temperature gradient) than the original model.

\begin{figure}[!htb]
\centering{\includegraphics{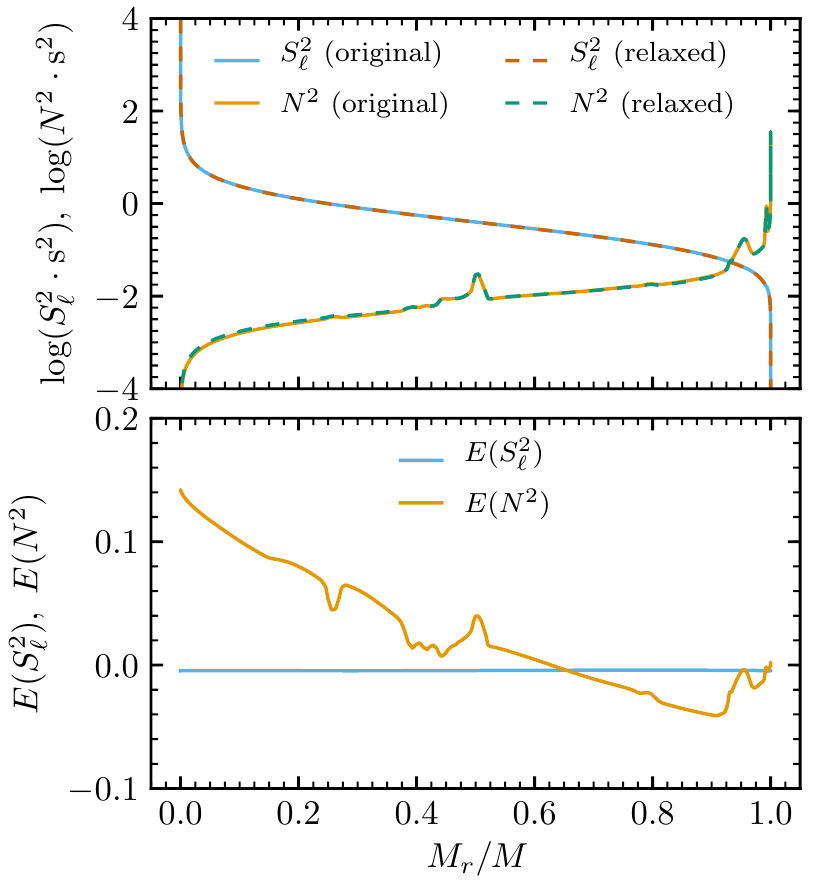}}
\caption{Comparison of the square of the Lamb ($S_{\ell}^{2})$ and
  \BV\ ($N^{2}$) frequencies for $\ell=1$ modes of the original and
  relaxed \MESA\ models.  The upper panel plots these data as a
  function of fractional mass $M_{r}/M$. The lower panel shows
  the relative differences, defined as
  $E(N^{2}) = (N^{2}_{\rm relax} - N^{2}_{\rm orig})/N^{2}_{\rm orig}$, and similarly for $S_{\ell}^{2}$.}
\label{fig:prop}
\end{figure}

\begin{figure}[!htb]
\centering{\includegraphics{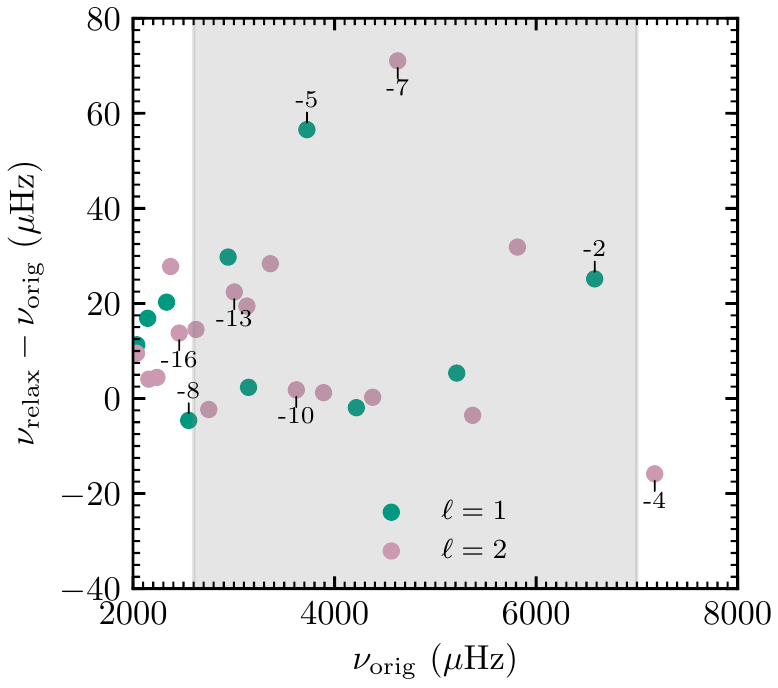}}
\caption{Differences between the adiabatic frequencies $\nu$ of $\ell=1$ and
  $\ell=2$ g-modes in the relaxed and original \mesa\ models. Selected
  modes are labeled by their radial order $\tilde{n}$. The shaded
  region marks the range of frequencies seen in the {\it Kepler}
  observations of \kic.}
\label{fig:delta_nu}
\end{figure}

\begin{figure*}[!htb]
\centering{\includegraphics{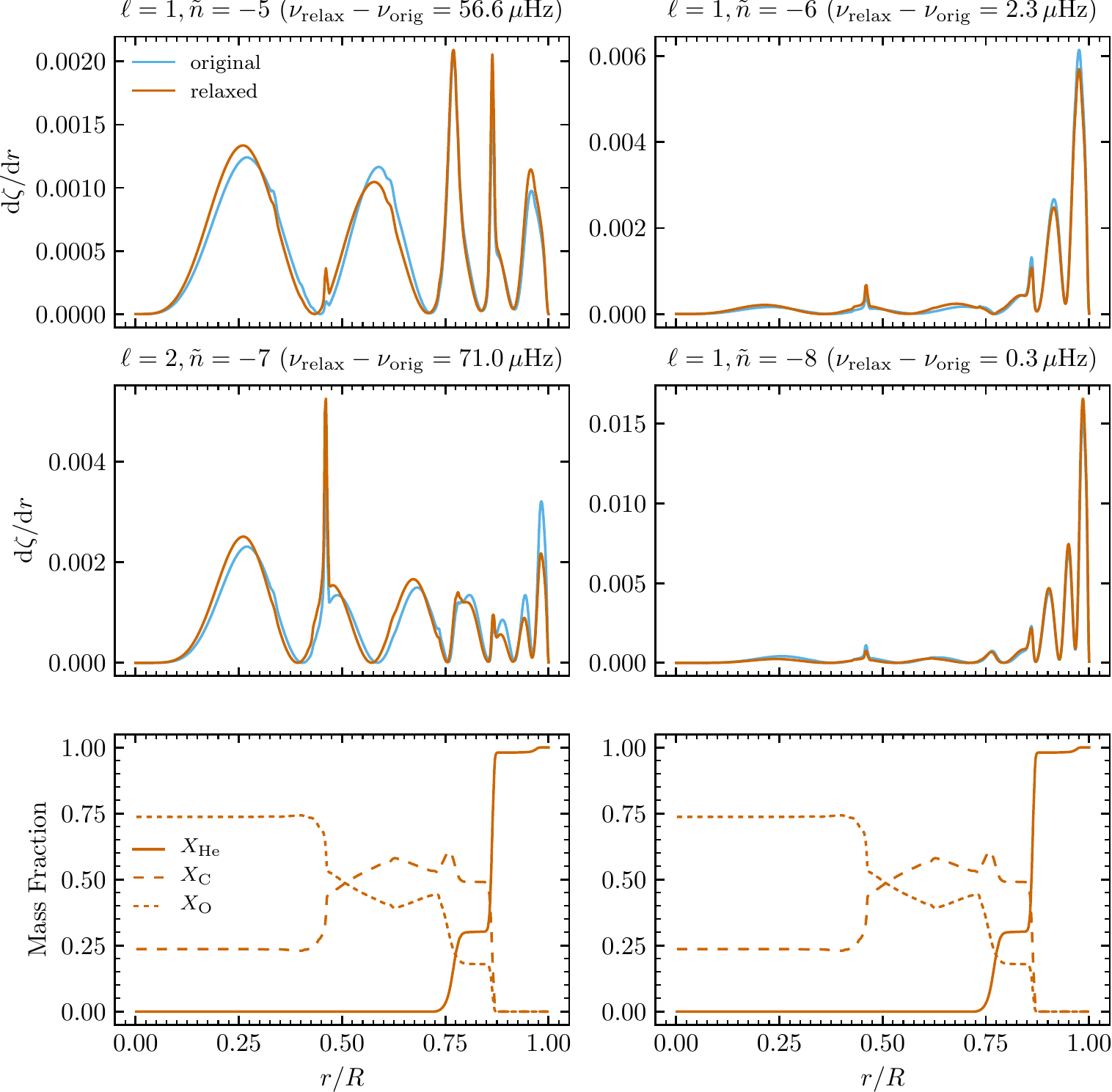}}
\caption{Weight functions for four adiabatic modes of the original and relaxed models;
a pair of adjacent radial order $\ell$=1 modes (top row) and a
pair of adjacent radial order $\ell$=2 modes (middle row).
The original and relaxed models share the same mass fraction profile (bottom row).
}
\label{fig:weight}
\end{figure*}

\section{Frequency Differences}
\label{sec:pulse}

Figure~\ref{fig:prop} shows the propagation diagram
\citep[e.g.,][]{Unno:1989aa} for dipole ($\ell=1$) modes of the
original and relaxed models. The upper panel plots the square of the
Lamb and \BV\ frequencies as a function of fractional mass $M_{r}/M$,
for the two models. The lower panel shows the relative difference
between these critical frequencies. While the models exhibit almost
identical $S_{\ell}^{2}$, the relaxed model has a larger $N^{2}$ in
its core than the original model by up to $\approx 14\%$; while
$N^{2}$ is smaller in the surface layers by up to $\approx 4\%$.

Assuming the magnitude of the temperature gradient in the interior 
to be much smaller than the adiabatic gradient, the \BV\ frequency reduces to
\citep[e.g.,][]{cox_1980_aa,bildsten_1995_aa,aerts_2010_aa}
\begin{equation}
  N^2 = \frac{1}{3 \Gamma_1 \chir H^2} \frac{k_B T}{\mu_I m_p}
\ ,
\label{eq:bv_simple}
\end{equation}
where 
$\Gamma_1$ is the first adiabatic index, 
$\chi_{\rho}$ is the density exponent $[\partial({\rm ln}P)/\partial({\rm ln}T)]_{\rho,\mu_I}$,
$H$ is the pressure scale height,
$k_B$ is the Boltzmann constant,
$m_p$ is the mass of the proton, and 
$\mu_I$ is the mean molecular weight of the ions.
There is a degeneracy between the $T$ and $\mu_I$ profiles in their effect on the \BV\ frequency,
$N^2 \propto T / \mu_I$. An inaccurate interior temperature profile can then directly impact inferences 
about the interior composition profile.

The \BV\ frequency differences in Figure~\ref{fig:prop} translate into corresponding differences 
in the g-mode frequencies.
We demonstrate this in Fig.~\ref{fig:delta_nu}, which
plots the difference between the $\ell=1$ and $\ell=2$ g-mode
frequencies of the relaxed model ($\nu_{\rm relax}$) and the original
model ($\nu_{\rm orig}$), as a function of $\nu_{\rm
  orig}$. Frequencies are calculated using release 5.2 of the
\GYRE\ software instrument \citep{townsend_2013_aa,townsend_2018_aa},
and selected modes are labeled by their radial order $\tilde{n}$ in
the \citet{Takata:2006ab} extension to the standard
Eckart-Osaki-Scuflaire classification scheme described e.g. by
\citet{Unno:1989aa}. The figure reveals frequency differences ranging
from $\approx -20\,{\rm \mu Hz}$ up to $\approx 70\,{\rm \mu
  Hz}$. There is no obvious pattern to the frequency differences, from
one mode to the next, although the scatter appears to reduce toward
larger values of $|\tilde{n}|$.

The frequency differences demonstrated here can be regarded as a
measure of the error introduced by assuming $L_{r} \propto M_{r}$
during seismic modeling. Therefore, although
\citet{giammichele_2018_aa} report that their model frequencies match
the observed frequencies to better than $\approx 0.6\,{\rm nHz}$, the
true error will likely be orders of magnitude larger --- and may have
an impact on the conclusions regarding core mass, radius, composition
etc. that they draw from their modeling.

Figure \ref{fig:weight} compares the weight functions 
of the original and relaxed models for 
pairs of adjacent radial order $\ell=1$ and $\ell~=~2$ adiabatic modes
The two pairs are chosen so that 
one mode shows a large frequency difference between the original and relaxed
models, but its radial order neighbor shows a small frequency difference.
Following \citet{kawaler_1985_aa}, the  weight function is
 \begin{equation}
  \frac{{\rm d}\zeta}{{\rm d}r} = \frac{[C({\bf y},r) + N({\bf y},r) + G({\bf y},r)] \rho r^2 }{\int_{r=0}^{r=R} T({\bf y},r) \rho r^2 {\rm d}r}
\ ,
\label{eq:dzdr}
\end{equation}
where 
$C({\bf y},r)$ contains the Lamb frequency,
$N({\bf y},r)$ varies with the \BV\ frequency,
$G({\bf y},r)$ involves the gravitational eigenfunctions,
$T({\bf y},r)$ is proportional to the kinetic energy density,
and ${\bf y} = (y_1,y_2,y_3,y_4)$ are the \citet{dziembowski_1971_aa} variables.
The frequency of an adiabatic mode is then 
\begin{equation}
\nu^2 =  \zeta = \int_{r=0}^{r=R} \frac{{\rm d}\zeta}{{\rm d}r} \cdot {\rm d}r
\ .
\label{eq:work}
\end{equation}

The weight function for the original and relaxed models is dominated by 
the $N({\bf y},r)$ term except for the surface layers,
and the change in the weight function due to a change in $N^2$ is
\begin{equation}
  \delta \left (\frac{{\rm d}\zeta}{{\rm d}r} \right )  = \frac{N({\bf y},r) (\delta N^2 / N^2) \rho r^2}{T(R)}
=  \frac{{\rm d}\zeta}{{\rm d}r} \frac{\delta N^2}{N^2}
\ .
\label{eq:delta}
\end{equation}
That is, the change in the weight function in going from the original
to the relaxed model is given by the weight function of the original
model times the fractional change in $N^2$.  The lower panel of
Figure~\ref{fig:prop} shows $\delta N^2/N^2$ is positive for 
$M_r/M \lesssim$\,0.6 and negative by a smaller amount for 
$M_r/M \gtrsim$\,0.6, again ignoring the surface layers. 
$M_r/M \simeq$\,0.6 corresponds to the location where the 
C-O profiles cross.
Equation~\ref{eq:delta} predicts these changes in $N^2$ will
increase the weight function in the inner region ($M_r/M
\lesssim$\,0.6) of the relaxed model, and decrease the weight function
in the outer region ($M_r/M \gtrsim$\,0.6) of the relaxed model by a
smaller amount.

The left column of Figure~\ref{fig:weight} verifies the expectations
from Equation~\ref{eq:delta}. The C-O crossover occurs at $r/R
\simeq$\,0.5.  The amplitude of the weight functions at the 
crossover is relatively small for the $\ell=1, \tilde{n}=-5$ 
mode but larger for the $\ell\,=\,2, \tilde{n}=-7$ mode.
Below the crossover, the amplitude of the weight function in the
relaxed model is larger than in the original model. Above the
crossover, the amplitude in the relaxed model is smaller.
In addition, the change in the amplitude
below the crossover is larger than the change in the amplitude above the crossover. 
The net effect of the weight
function redistribution on either side of the crossover causes a
shift in $\zeta$, the area under the weight function curves, towards
larger mode frequencies in the relaxed model 
(56.6~$\mu$Hz for the $\ell=1,\tilde{n}=-5$ mode
and 71.0~$\mu$Hz for the $\ell=2,\tilde{n}=-7$ mode).

The right column of Figure~\ref{fig:weight} shows the same 
behavior; larger amplitudes below the C-O crossover, smaller
amplitude decreases above the crossover.  However, the weight functions
are concentrated toward the surface layers where $N^2$ does not
significantly change (see Figure~\ref{fig:prop}). Hence, these mode
frequencies are relatively unaffected in transitioning from the original to the
relaxed model
(2.3~$\mu$Hz for the $\ell=1,\tilde{n}=-6$ mode
and 0.3 $\mu$Hz for the $\ell=2,\tilde{n}=-8$ mode).

Figures \ref{fig:prop} and \ref{fig:weight} encapsulate two additional messages. 
First, the modes which best probe the interior, those whose weight
functions are large in the interior, are also the modes most affected
by the change in the thermal structure in transitioning from the 
$L_r \not\propto M_r$ 
original 
model 
to the 
$L_r \propto M_r$ 
relaxed model.
Second, we do not find a large switch between a mode being confined to the core 
to being confined to the envelope when transitioning from the original to the relaxed model. 
That is, the physics changing the mode frequencies is the 
thermal profile rather than mode trapping.

\begin{figure}[!htb]
\centering{\includegraphics[width=1.0\columnwidth]{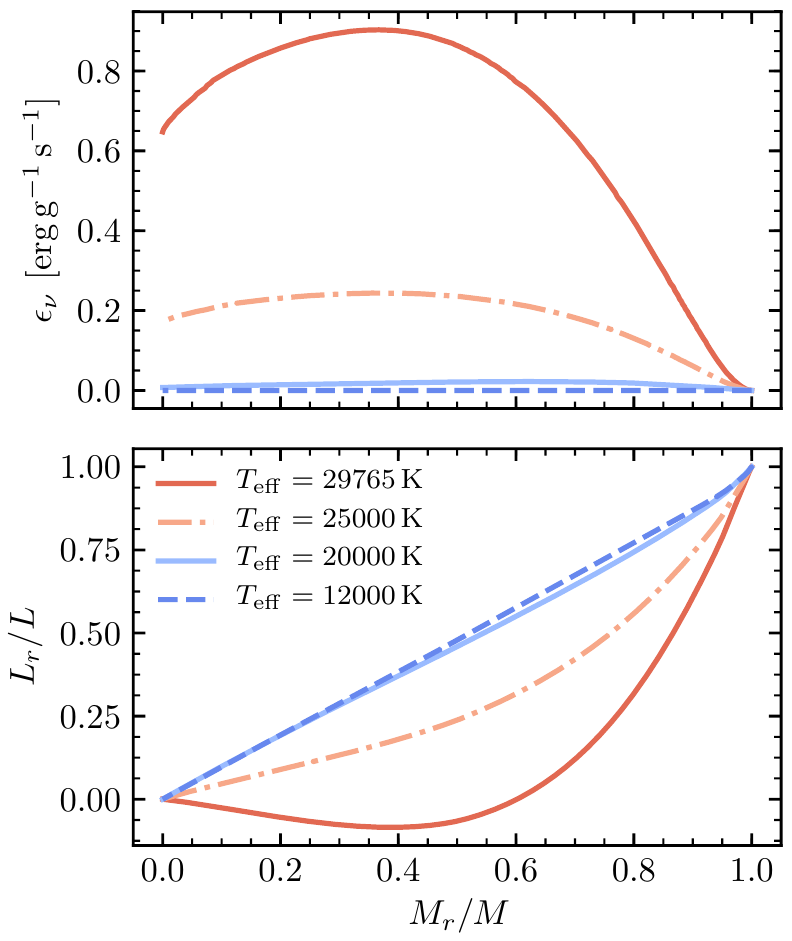}}
\caption{Profiles for the original model as the WD evolves toward
  cooler $T_{\rm eff}$. The upper panel shows the neutrino cooling
  rate, which dominates the interior regions of the WD for the $T_{\rm
    eff}= 29,765\,{\rm K}$ curve, yielding $L_r \not\propto M_r$.
  The luminosity profiles in the lower panel show that $L_r \propto
  M_r$ does not occur until $T_{\rm eff} \lesssim 20,000\,{\rm K}$.  }
\label{fig:Levol}
\end{figure}

\section{Cooling of the Evolution Model}
\label{sec:cooling}

When does $L_r \propto M_r$ hold in our original model?
Figure~\ref{fig:Levol} shows the evolution of the original
model as the WD cools down.  Plasmon neutrino emission dominates
the energy loss budget for average-mass CO WDs with $T_{\rm eff}
\gtrsim 25,000\,{\rm K}$ \citep[e.g.,][]{vila_1966_aa, kutter_1969_aa,
  winget_2004_aa, bischoff-kim_2018_aa}.  The lower end of this range
overlaps the observed $T_{\rm eff}$ for the DBV and V777 Her classes,
of which \kic is a member.  Neutrino losses explain why the 
$T_{\rm eff}= 29,765\,{\rm K}$ luminosity profile is negative 
(inward heat flux)
from the center to $M_r/M \simeq$\,0.6.

As the evolution of the original model continues, photons leaving the
WD surface begin to dominate the cooling as the electrons transition
to a strongly degenerate plasma \citep{van-horn_1971_aa}.  
Energy transport
in the interior is dominated by conduction, driven primarily by
electron-ion scattering. Energy transport in the outer layers is
dominated by radiation or convection associated with the partial
ionization of the most abundant element near the surface
\citep[e.g.,][]{winget_2008_aa,althaus_2010_aa}.
For DBV WDs, the partial
ionization of He occurs around $T_{\rm eff} \simeq 30,000 \, {\rm K}$, 
leading to convection and pulsations in relatively hot WDs.
Figure~\ref{fig:Levol} shows the relation $L_r \propto M_r$ is
approximately satisfied in the interior of our CO WD model
only after $T_{\rm eff} \lesssim 20,000\,{\rm K}$.
These lower temperatures are associated with DAV WDs, where partial
ionization in their hydrogen atmospheres leads to the onset of
convection and pulsations.
This suggests that the approximation $L_r \propto M_r$ may be more
reliable for asteroseismic studies of this cooler class of objects.

\section{Conclusions}
\label{sec:conclude}

We have generated a pair of WD models that are `spectroscopic twins',
having the same effective temperature, surface gravity, and
abundances.
One model has an enforced $L_r \propto M_r$ luminosity distribution
and the other model has a luminosity distribution given by the star's
previous evolution.  The low-order g-mode oscillation frequencies of
the two models differ by up to $\simeq 70\,{\rm \mu Hz}$, but in an
uneven manner.

This result suggests that by neglecting the proper thermal structure
of the star (e.g., accounting for the effect of plasmon neutrino
losses), the model frequencies calculated by using an $L_r \propto M_r$ 
model may have uncorrected random errors as large as $\simeq 70\,{\rm \mu Hz}$.  
To aid interpretation of this frequency difference, Table 9
of \citet{giammichele_2017_ab} 
lists the uncertainty in the derived WD mass, radius, 
and central oxygen mass fraction for frequency differences of 
0.001, 0.01, and $10\,{\rm \mu Hz}$.
For example, a $10\,{\rm \mu Hz}$
frequency difference translates into an uncertainty in the fit precision of 
$\simeq 4\%$ in the WD mass,
$\simeq 3\%$ in the WD radius, and 
$\simeq 1\%$ in the central $^{16}$O mass fraction.
Their Table 9 shows a factor of 10 increase in the frequency
difference causes about an order of magnitude increase in the uncertainty of the derived WD properties,
with a fitting trend of larger masses, smaller radii, and smaller $^{16}$O mass fractions for larger frequency differences.
Assuming this trend holds for larger frequency differences, 
then a linear extrapolation to a mean frequency difference of $30\,{\rm \mu Hz}$
betwen the original $L_r \not\propto M_r$  and relaxed $L_r \propto M_r$ models
suggests an uncertainty in the template model fit precision of 
$\simeq 12\%$ in WD mass, 
$\simeq 9\%$ in the WD radius, and 
$\simeq 3\%$ in the central $^{16}$O mass fraction. 
Alternatively, a frequency difference of $60\,{\rm \mu Hz}$ for modes that are 
especially sensitive to the core composition translates to a $\simeq 6\%$ uncertainty 
in the central $^{16}$O mass fraction.
Finally, Figure~\ref{fig:structure} shows a 10-20\% difference in the core temperature 
between the original $L_r \not\propto M_r$  and relaxed $L_r \propto M_r$ models.
If the scaling of Equation~\ref{eq:bv_simple} is roughly correct, there may be a 10-20\% difference 
in the derived $\mu_I$, which may translate into larger uncertainties in the 
C/O mass fractions.

We encourage future generations of $T_{\rm eff}
\gtrsim 20,000\,{\rm K}$ WD template models to consider using
luminosity profiles informed by evolution models 
or to include a luminosity profile as part of the template model
fitting process.

\acknowledgements

We thank Noemi Giammichele for discussions and generously sharing
detailed WD template model profiles, Steve Kawaler for communications and 
insights, and the anonymous referee for suggestions that improved this Letter.
This project was supported by NSF under the 
SI2
program (ACI-1663684, ACI-1663688, ACI-1663696), 
the 
AAG 
program (AST-1716436) and grant 
PHY-1430152 for the 
PFC 
``Joint Institute for Nuclear Astrophysics
- Center for the Evolution of the Elements'' (JINA-CEE).  
C.E.F. was supported by a Ford Foundation Predoctoral Fellowship, an NSF GRFP
Fellowship (DGE1424871), and a Edward J. Petry Graduate Fellowship from MSU.  
A.T. is supported as a Research Associate at the Belgian
Scientific Research Fund (F.R.S-FNRS).  This research made extensive
use of the SAO/NASA Astrophysics Data System (ADS).

\software{\MESA\ \citep[][\url{http://mesa.sourceforge.net}]{paxton_2011_aa,paxton_2013_aa,paxton_2015_aa,paxton_2018_aa},
\GYRE\ \citep[][\url{https://bitbucket.org/rhdtownsend/gyre/wiki/Home}]{townsend_2013_aa,townsend_2018_aa}
}

\bibliographystyle{aasjournal}

\begin{thebibliography}{}
\expandafter\ifx\csname natexlab\endcsname\relax\def\natexlab#1{#1}\fi
\providecommand{\url}[1]{\href{#1}{#1}}
\providecommand{\dodoi}[1]{doi:~\href{http://doi.org/#1}{\nolinkurl{#1}}}
\providecommand{\doeprint}[1]{\href{http://ascl.net/#1}{\nolinkurl{http://ascl.net/#1}}}
\providecommand{\doarXiv}[1]{\href{https://arxiv.org/abs/#1}{\nolinkurl{https://arxiv.org/abs/#1}}}

\bibitem[{{Aerts} {et~al.}(2010){Aerts}, {Christensen-Dalsgaard}, \&
  {Kurtz}}]{aerts_2010_aa}
{Aerts}, C., {Christensen-Dalsgaard}, J., \& {Kurtz}, D.~W. 2010,
  {Asteroseismology} ({Springer Science+Business Media B.V.})

\bibitem[{{Althaus} {et~al.}(2010){Althaus}, {Córsico}, {Isern}, \&
  {Garc{\'{\i}}a-Berro}}]{althaus_2010_aa}
{Althaus}, L.~G., {Córsico}, A.~H., {Isern}, J., \& {Garc{\'{\i}}a-Berro}, E.
  2010, \aapr, 18, 471, \dodoi{10.1007/s00159-010-0033-1}

\bibitem[{{Bildsten} \& {Cutler}(1995)}]{bildsten_1995_aa}
{Bildsten}, L., \& {Cutler}, C. 1995, \apj, 449, 800, \dodoi{10.1086/176099}

\bibitem[{{Bischoff-Kim} \& {Metcalfe}(2011)}]{bischoff-kim_2011_aa}
{Bischoff-Kim}, A., \& {Metcalfe}, T.~S. 2011, \mnras, 414, 404,
  \dodoi{10.1111/j.1365-2966.2011.18396.x}

\bibitem[{{Bischoff-Kim} \& {Montgomery}(2018)}]{bischoff-kim_2018_aa}
{Bischoff-Kim}, A., \& {Montgomery}, M.~H. 2018, \aj, 155, 187,
  \dodoi{10.3847/1538-3881/aab70e}

\bibitem[{{Bischoff-Kim} {et~al.}(2014){Bischoff-Kim}, {Østensen}, {Hermes},
  \& {Provencal}}]{bischoff-kim_2014_aa}
{Bischoff-Kim}, A., {Østensen}, R.~H., {Hermes}, J.~J., \& {Provencal}, J.~L.
  2014, \apj, 794, 39, \dodoi{10.1088/0004-637X/794/1/39}

\bibitem[{{Brassard} {et~al.}(1992){Brassard}, {Fontaine}, {Wesemael}, \&
  {Hansen}}]{brassard_1992_aa}
{Brassard}, P., {Fontaine}, G., {Wesemael}, F., \& {Hansen}, C.~J. 1992, \apjs,
  80, 369, \dodoi{10.1086/191668}

\bibitem[{{Cox}(1980)}]{cox_1980_aa}
{Cox}, J.~P. 1980, {Theory of stellar pulsation}, {Princeton Series in
  Astrophysics} ({Princeton: University Press})

\bibitem[{{De Ger{\'o}nimo} {et~al.}(2017){De Ger{\'o}nimo}, {Althaus},
  {C{\'o}rsico}, {Romero}, \& {Kepler}}]{Geronimo17}
{De Ger{\'o}nimo}, F.~C., {Althaus}, L.~G., {C{\'o}rsico}, A.~H., {Romero},
  A.~D., \& {Kepler}, S.~O. 2017, \aap, 599, A21,
  \dodoi{10.1051/0004-6361/201629806}

\bibitem[{{De Ger{\'o}nimo} {et~al.}(2018){De Ger{\'o}nimo}, {Althaus},
  {C{\'o}rsico}, {Romero}, \& {Kepler}}]{Geronimo18}
---. 2018, \aap, 613, A46, \dodoi{10.1051/0004-6361/201731982}

\bibitem[{{Dziembowski}(1971)}]{dziembowski_1971_aa}
{Dziembowski}, W.~A. 1971, \actaa, 21, 289

\bibitem[{{Farmer} {et~al.}(2016){Farmer}, {Fields}, {Petermann}, {Dessart},
  {Cantiello}, {Paxton}, \& {Timmes}}]{farmer_2016_aa}
{Farmer}, R., {Fields}, C.~E., {Petermann}, I., {et~al.} 2016, \apjs, 227, 22,
  \dodoi{10.3847/1538-4365/227/2/22}

\bibitem[{{Fields} {et~al.}(2016){Fields}, {Farmer}, {Petermann}, {Iliadis}, \&
  {Timmes}}]{fields_2016_aa}
{Fields}, C.~E., {Farmer}, R., {Petermann}, I., {Iliadis}, C., \& {Timmes},
  F.~X. 2016, \apj, 823, 46, \dodoi{10.3847/0004-637X/823/1/46}

\bibitem[{{Fields} {et~al.}(2018){Fields}, {Timmes}, {Farmer}, {Petermann},
  {Wolf}, \& {Couch}}]{fields_2018_aa}
{Fields}, C.~E., {Timmes}, F.~X., {Farmer}, R., {et~al.} 2018, \apjs, 234, 19,
  \dodoi{10.3847/1538-4365/aaa29b}

\bibitem[{{Fontaine} \& {Brassard}(2008)}]{fontaine_2008_aa}
{Fontaine}, G., \& {Brassard}, P. 2008, \pasp, 120, 1043,
  \dodoi{10.1086/592788}

\bibitem[{{Fontaine} {et~al.}(2001){Fontaine}, {Brassard}, \&
  {Bergeron}}]{fontaine_2001_aa}
{Fontaine}, G., {Brassard}, P., \& {Bergeron}, P. 2001, \pasp, 113, 409,
  \dodoi{10.1086/319535}

\bibitem[{{Giammichele} {et~al.}(2017{\natexlab{a}}){Giammichele}, {Charpinet},
  {Brassard}, \& {Fontaine}}]{giammichele_2017_aa}
{Giammichele}, N., {Charpinet}, S., {Brassard}, P., \& {Fontaine}, G.
  2017{\natexlab{a}}, \aap, 598, A109, \dodoi{10.1051/0004-6361/201629935}

\bibitem[{{Giammichele} {et~al.}(2017{\natexlab{b}}){Giammichele}, {Charpinet},
  {Fontaine}, \& {Brassard}}]{giammichele_2017_ab}
{Giammichele}, N., {Charpinet}, S., {Fontaine}, G., \& {Brassard}, P.
  2017{\natexlab{b}}, \apj, 834, 136, \dodoi{10.3847/1538-4357/834/2/136}

\bibitem[{{Giammichele} {et~al.}(2016){Giammichele}, {Fontaine}, {Brassard}, \&
  {Charpinet}}]{giammichele_2016_aa}
{Giammichele}, N., {Fontaine}, G., {Brassard}, P., \& {Charpinet}, S. 2016,
  \apjs, 223, 10, \dodoi{10.3847/0067-0049/223/1/10}

\bibitem[{{Giammichele} {et~al.}(2018){Giammichele}, {Charpinet}, {Fontaine},
  {Brassard}, {Green}, {Van Grootel}, {Bergeron}, {Zong}, \&
  {Dupret}}]{giammichele_2018_aa}
{Giammichele}, N., {Charpinet}, S., {Fontaine}, G., {et~al.} 2018, \nat, 554,
  73, \dodoi{10.1038/nature25136}

\bibitem[{{Hansen}(2004)}]{hansen_2004_aa}
{Hansen}, B. 2004, \physrep, 399, 1, \dodoi{10.1016/j.physrep.2004.07.001}

\bibitem[{{Kawaler} {et~al.}(1985){Kawaler}, {Winget}, \&
  {Hansen}}]{kawaler_1985_aa}
{Kawaler}, S.~D., {Winget}, D.~E., \& {Hansen}, C.~J. 1985, \apj, 295, 547,
  \dodoi{10.1086/163398}

\bibitem[{{Kutter} \& {Savedoff}(1969)}]{kutter_1969_aa}
{Kutter}, G.~S., \& {Savedoff}, M.~P. 1969, \apj, 156, 1021,
  \dodoi{10.1086/150033}

\bibitem[{{Liebert}(1980)}]{liebert_1980_aa}
{Liebert}, J. 1980, \araa, 18, 363, \dodoi{10.1146/annurev.aa.18.090180.002051}

\bibitem[{{Maschberger}(2013)}]{maschberger_2013_aa}
{Maschberger}, T. 2013, \mnras, 429, 1725, \dodoi{10.1093/mnras/sts479}

\bibitem[{{Metcalfe}(2005)}]{metcalfe_2005_aa}
{Metcalfe}, T.~S. 2005, \mnras, 363, L86,
  \dodoi{10.1111/j.1745-3933.2005.00091.x}

\bibitem[{{Metcalfe} {et~al.}(2002){Metcalfe}, {Salaris}, \&
  {Winget}}]{metcalfe_2002_aa}
{Metcalfe}, T.~S., {Salaris}, M., \& {Winget}, D.~E. 2002, \apj, 573, 803,
  \dodoi{10.1086/340796}

\bibitem[{{{\O}stensen} {et~al.}(2011){{\O}stensen}, {Bloemen},
  {Vu\v{c}kovi\'{c}}, {Aerts}, {Oreiro}, {Kinemuchi}, {Still}, \&
  {Koester}}]{ostensen_2011_aa}
{{\O}stensen}, R.~H., {Bloemen}, S., {Vu\v{c}kovi\'{c}}, M., {et~al.} 2011,
  \apjl, 736, L39, \dodoi{10.1088/2041-8205/736/2/L39}

\bibitem[{{Paxton} {et~al.}(2011){Paxton}, {Bildsten}, {Dotter}, {Herwig},
  {Lesaffre}, \& {Timmes}}]{paxton_2011_aa}
{Paxton}, B., {Bildsten}, L., {Dotter}, A., {et~al.} 2011, \apjs, 192

\bibitem[{{Paxton} {et~al.}(2013){Paxton}, {Cantiello}, {Arras}, {Bildsten},
  {Brown}, {Dotter}, {Mankovich}, {Montgomery}, {Stello}, {Timmes}, \&
  {Townsend}}]{paxton_2013_aa}
{Paxton}, B., {Cantiello}, M., {Arras}, P., {et~al.} 2013, \apjs, 208

\bibitem[{{Paxton} {et~al.}(2015){Paxton}, {Marchant}, {Schwab}, {Bauer},
  {Bildsten}, {Cantiello}, {Dessart}, {Farmer}, {Hu}, {Langer}, {Townsend},
  {Townsley}, \& {Timmes}}]{paxton_2015_aa}
{Paxton}, B., {Marchant}, P., {Schwab}, J., {et~al.} 2015, \apjs, 220, 15,
  \dodoi{10.1088/0067-0049/220/1/15}

\bibitem[{{Paxton} {et~al.}(2018){Paxton}, {Schwab}, {Bauer}, {Bildsten},
  {Blinnikov}, {Duffell}, {Farmer}, {Goldberg}, {Marchant}, {Sorokina},
  {Thoul}, {Townsend}, \& {Timmes}}]{paxton_2018_aa}
{Paxton}, B., {Schwab}, J., {Bauer}, E.~B., {et~al.} 2018, \apjs, 234, 34,
  \dodoi{10.3847/1538-4365/aaa5a8}

\bibitem[{{Romero} {et~al.}(2012){Romero}, {Córsico}, {Althaus}, {Kepler},
  {Castanheira}, \& {Miller Bertolami}}]{romero_2012_aa}
{Romero}, A.~D., {Córsico}, A.~H., {Althaus}, L.~G., {et~al.} 2012, \mnras,
  420, 1462, \dodoi{10.1111/j.1365-2966.2011.20134.x}

\bibitem[{{Romero} {et~al.}(2017){Romero}, {Córsico}, {Castanheira}, {De
  Ger{}\'onimo}, {Kepler}, {Koester}, {Kawka}, {Althaus}, {Hermes}, {Bonato},
  \& {Gianninas}}]{romero_2017_aa}
{Romero}, A.~D., {Córsico}, A.~H., {Castanheira}, B.~G., {et~al.} 2017, \apj,
  851, 60, \dodoi{10.3847/1538-4357/aa9899}

\bibitem[{{Salpeter}(1955)}]{salpeter_1955_aa}
{Salpeter}, E.~E. 1955, \apj, 121, 161, \dodoi{10.1086/145971}

\bibitem[{{Scalo}(1986)}]{scalo_1986_aa}
{Scalo}, J.~M. 1986, \fcp, 11, 1

\bibitem[{{Takata}(2006)}]{Takata:2006ab}
{Takata}, M. 2006, \pasj, 58, 893, \dodoi{10.1093/pasj/58.5.893}

\bibitem[{{Townsend} {et~al.}(2018){Townsend}, {Goldstein}, \&
  {Zweibel}}]{townsend_2018_aa}
{Townsend}, R.~H.~D., {Goldstein}, J., \& {Zweibel}, E.~G. 2018, \mnras, 475,
  879, \dodoi{10.1093/mnras/stx3142}

\bibitem[{{Townsend} \& {Teitler}(2013)}]{townsend_2013_aa}
{Townsend}, R.~H.~D., \& {Teitler}, S.~A. 2013, \mnras, 435, 3406,
  \dodoi{10.1093/mnras/stt1533}

\bibitem[{{Unno} {et~al.}(1989){Unno}, {Osaki}, {Ando}, {Saio}, \&
  {Shibahashi}}]{Unno:1989aa}
{Unno}, W., {Osaki}, Y., {Ando}, H., {Saio}, H., \& {Shibahashi}, H. 1989,
  {Nonradial oscillations of stars} (University of Tokyo Press, Tokyo)

\bibitem[{{van Horn}(1971)}]{van-horn_1971_aa}
{van Horn}, H.~M. 1971, in IAU Symposium, Vol.~42, White Dwarfs, ed. W.~J.
  {Luyten}, 97

\bibitem[{{Vila}(1966)}]{vila_1966_aa}
{Vila}, S.~C. 1966, \apj, 146, 437, \dodoi{10.1086/148908}

\bibitem[{{Winget} \& {Kepler}(2008)}]{winget_2008_aa}
{Winget}, D.~E., \& {Kepler}, S.~O. 2008, \araa, 46, 157,
  \dodoi{10.1146/annurev.astro.46.060407.145250}

\bibitem[{{Winget} {et~al.}(2004){Winget}, {Sullivan}, {Metcalfe}, {Kawaler},
  \& {Montgomery}}]{winget_2004_aa}
{Winget}, D.~E., {Sullivan}, D.~J., {Metcalfe}, T.~S., {Kawaler}, S.~D., \&
  {Montgomery}, M.~H. 2004, \apjl, 602, L109, \dodoi{10.1086/382591}

\bibitem[{{Winget} {et~al.}(1982){Winget}, {van Horn}, {Tassoul}, {Fontaine},
  {Hansen}, \& {Carroll}}]{winget_1982_aa}
{Winget}, D.~E., {van Horn}, H.~M., {Tassoul}, M., {et~al.} 1982, \apjl, 252,
  L65, \dodoi{10.1086/183721}

\bibitem[{{Zong} {et~al.}(2016){Zong}, {Charpinet}, {Vauclair}, {Giammichele},
  \& {Van Grootel}}]{zong_2016_aa}
{Zong}, W., {Charpinet}, S., {Vauclair}, G., {Giammichele}, N., \& {Van
  Grootel}, V. 2016, \aap, 585, A22, \dodoi{10.1051/0004-6361/201526300}

\end{thebibliography}


\end{document}